\documentclass[12pt]{article}
\usepackage{float}
\usepackage{color}
\usepackage{fancybox,amsmath,amssymb,rotating}

\def\bc        {\begin{center}}
\def\ec        {\end{center}}

\definecolor{black}{cmyk}{1,0,1.0,1.0}
\definecolor{grey0}{rgb}{0.7,0.7,0.7}
\definecolor{grey1}{rgb}{0.6,0.6,0.6}
\definecolor{grey2}{rgb}{0.4,0.4,0.4}
\definecolor{grey3}{rgb}{0.2,0.2,0.2}
\definecolor{lightyellow}{cmyk}{0,0,0.5,0}
\definecolor{lightred}{rgb}{1,0.5,0.5}
\definecolor{lightgreen}{rgb}{0.5,1,0.5}
\definecolor{lightblue}{rgb}{0.5,0.5,1}
\definecolor{darkred}{rgb}{0.8,0,0}
\definecolor{darkgreen}{rgb}{0,0.4,0}
\definecolor{darkcyan}{cmyk}{1,0.3,0.3,0.3}
\definecolor{darkblue}{rgb}{0,0,0.6}
\definecolor{lightbrown}{rgb}{0.7,0.3,0.3}
\definecolor{darkbrown}{rgb}{0.5,0,0}
\definecolor{bluegreen}{rgb}{0,0.5,0.5}

\begin{document}
\bc
{\bf\Large Hadron Spectroscopy, Results and Ideas.}\\[6ex]Eberhard
Klempt\\
HISKP, Bonn University, Nussallee 14-16\\[2ex]

{\footnotesize Invited Talk presented at the International Workshop
 on\\[0.5ex]
NEW PARTIAL WAVE ANALYSIS TOOLS FOR NEXT GENERATION \\HADRON
SPECTROSCOPY EXPERIMENTS\\ June 20-22, 2012 Camogli,
Italy}\vspace{4mm}\ec \subsection*{New Results from the BnGa PWA}

The BnGa partial wave analysis group has fitted a very large set of
data in a coupled-channel analysis \cite{Anisovich:2011fc}. The data
comprise all major reactions like $\gamma p\to p\pi^0, n\pi^+,
p\eta, \Lambda K^+, \Sigma^0 K^+, \Sigma^+ K^0$, and $p\pi^0\pi^0$
and is further constrained by pion induced inelastic reactions like
$\pi N\to \Lambda K, \Sigma K$, $N\eta$, $N\pi\pi$, and by the real
and imaginary part of $\pi N$ scattering amplitudes, alternative
from Karlsruhe-Helsinki \cite{Hohler:1979yr} or from the GWU
\cite{Arndt:2006bf}. The analysis uses a K-matrix with up to eight
constrained channels ($N\pi$, $N\eta$, $\Lambda K$, $\Sigma K$,
$\Delta\pi$-low-$L$, $\Delta\pi$-high-$L$, $N(\pi\pi)_{\rm S-wave}$
and one channel ($N\omega$) which is not constrained by data and
which simulates unseen decay modes. Further decay modes like
$N(1520)3/2^-$, $N(1535)1/2^-$, $N(1680)5/2^+$, $N(1710)1/2^+$, or
$N\,f_2(1270)$ are added in a $D$-vector formalism which does not
take into account rescattering between the different channels. The
background is described by reggeized $t$-channel exchange of vector
mesons; for the reactions $\gamma p\to n\pi^+$ and $\gamma p\to
\Sigma^0 K^+$, $\pi$ and $K$ exchanges are admitted. $N$ and
$\Delta$ exchange in contribute via Born diagrams in the $s$ and the
$u$-channel. In each partial wave, non-resonant transitions like
$\gamma p\to p\eta$ are admitted. For most partial waves, these are
constants, in the wave with $J^P=1/2^-$, a simple function in $s$ is
used. Full account of the formalism is given in
\cite{Anisovich:2004zz}.

The results are summarized in Table~\ref{tab:bering}. A large number
of parameters on baryon resonances ($N_{\rm pp}$ in the Table) is
presented in \cite{Anisovich:2011fc}. In the Review of Particle
Properties \cite{PDG} there are now seven new resonances. The
one-star $N(1685)$ has been reported as narrow peak in analyses of
$\gamma n\to n\eta$ \cite{gra0}-\cite{Jaegle:2011sw}; it is proposed
to belong to an antidecuplet which would require $J^P=1/2^+$ quantum
numbers. $N(2040)3/2^+$ - with one-star - is seen in $J/\psi\to
p\bar p\pi$ \cite{bes}. Five resonances stem from the BnGa partial
wave analysis, even though most of these resonances have been
reported before but did not pass the threshold to a recognized
resonance. Four of the new resonances received a two-star status.
Early observations of $N(1875)3/2^-$, reported by Manley
\cite{Manley:1984jz}, Bell \cite{Bell:1983dm}, Cutkosky
\cite{Cutkosky:1980rh}, and Saxon \cite{Saxon:1979xu} had been
listed under $N(2080)$; they can now be found - jointly with the
BnGa result - under a new three-star $N(1875)3/2^-$. $N(1900)3/2^+$
was upgraded from two to three stars.
\begin{table}[pt]
\renewcommand{\arraystretch}{1.2}
{\footnotesize\bc
\begin{tabular}{lcllcllcl}
\hline\hline\\[-3ex] \quad Resonance&\hspace{-1.5mm}
Rating&\hspace{-2.5mm}$N_{\rm pp}$\hspace{-2.5mm}&\quad
Resonance&\hspace{-1.5mm} Rating&\hspace{-2.5mm}$N_{\rm pp}$&\quad
Resonance &\hspace{-1.5mm} Rating&\hspace{-2.5mm}$N_{\rm
pp}$\hspace{-2.5mm}\\[0.5ex]\hline $N(1440){1/2^+}$& **** &  13 & $N(1520){3/2^-}$& **** &  17                               &
$N(1535){1/2^-}$& **** &  15                               \\
$N(1650){1/2^-}$& **** &  18                               &
$N(1675){5/2^-}$& **** &  14                               &
$N(1680){5/2^+}$& **** &  17                              \\
$N(1685)$&*\cite{gra0}-\cite{Jaegle:2011sw} & & $N(1700){3/2^-}$&
***  &  15                               &
$N(1710){1/2^+}$& ***  &  14                                \\
$N(1720){3/2^+}$& **** &  17                               &
$N(1860){5/2^+}$& **\cite{Anisovich:2011fc}&\ 9 &
$N(1875){3/2^-}$ & ***\cite{Anisovich:2011fc}& 16  \\
$N(1880){1/2^+}$&**\cite{Anisovich:2011fc}&  20  & $N(1895){1/2^-}$
&**\cite{Anisovich:2011fc}& 17  &
$N(1900){3/2^+}$& ***\cite{Anisovich:2011fc}  &     18            \\
$N(1990){7/2^+}$& ** &  \ 9                                &
$N(2000){5/2^+}$& **   & 11                                &
$N(2040){3/2^+}$ & *\cite{bes}     &    \\
$N(2060){5/2^-}$&**\cite{Anisovich:2011fc}   & 13 & $N(2100){1/2^+}$
&*    &                  &
$N(2150){3/2^-}$   &**\cite{Anisovich:2011fc}& 11 \\
$N(2190){7/2^-}$&****        &          11                 &
$N(2220){7/2^-}$ &              ****   &\ 7                &
$N(2250){9/2^-}$ &**** &                                  \\
$N(2600){11/2^-}$ &***  &                                 &
$N(2700){13/2^+}$                &**  &                   \\
$\Delta(1232)$                    &**** &  \ 8                &
$\Delta(1600){3/2^+}$             &***  &  12              &
$\Delta(1620){1/2^-}$              &****&  10              \\
$\Delta(1700){3/2^-}$             &**** &   11              &
$\Delta(1750){1/2^+}$             &*    &                  &
$\Delta(1900){1/2^-}$             &**  &  13               \\
$\Delta(1905){5/2^+}$              &**** & 11                &
$\Delta(1910){1/2^+}$              &**** & 13                &
$\Delta(1920){3/2^+}$              &*** & 21               \\
$\Delta(1930){5/2^-}$           &***    &              &
$\Delta(1940){3/2^-}$             &*    & \ 5              &
$\Delta(1950){7/2^+}$              &****& 13                 \\
$\Delta(2000){5/2^+}$              &**  &                  &
$\Delta(2150){1/2^-}$              &*   &                &
$\Delta(2200){7/2^-}$               &*  &               \\
$\Delta(2300){9/2^+}$                &** &                &
$\Delta(2350){3/2^-}$               &*  &                 &
$\Delta(2390){7/2^+}$              &*    &              \\
$\Delta(2420){11/2^+}$              &****&                   &
$\Delta(2400){9/2^-}$               &****&                 &
$\Delta(2750){13/2^-}$                &**&               \\
$\Delta(2950){15/2^+}$             &**  &            \\[0.5ex]
\hline\hline
\end{tabular}
\renewcommand{\arraystretch}{1.0}
\end{center}}
\caption{\label{tab:bering}Nucleon and $\Delta$ resonances in the
new Review of Particle Properties~\cite{PDG}. In the BnGa analysis
$N_{\rm pp}$ particle properties were determined; 400 in total. Be
cautious, however, there are ambiguities. $N(1875)3/2^-$ and
$N(1900)3/2^+$ were promoted to three-star resonances.}
\end{table}

In the region above 1.8\,GeV, no unique solution was found by the
PWA. There were distinct solutions with two or three resonances with
$J^P=3/2^+$ or one or two resonances with $J^P=5/2^+$. Even for
these ``main" solutions, called BnGa2011-1 and BnGa2011-2, the
resulting amplitudes depend on the inclusion or exclusion of
high-mass resonances (above 2200\,MeV), on the inclusion or not of
additional channel couplings, also on start values. From the spread
of results within a class of solutions, error bars were derived. In
some cases different hypotheses lead to significantly different pole
positions, and the error bars become large. In the discussion, the
alternative values sometimes support different interpretations. Here
we use the values which support an interpretation under discussion,
either from BnGa2011-1 or from BnGa2011-2. Both are discussed in
\cite{Anisovich:2011ye}.
\subsection*{Interpretation}
\paragraph{Baryon from quark models and the lattice:}
The systematics of the baryon ground states were constitutive for
the development of quark models. In the harmonic oscillator (h.o.)
approximation, the quark model predicts a ladder of baryon
resonances with equidistant squared masses, alternating with
positive and negative parity, and this pattern survives
approximately in more realistic potentials (see, e.g.
\cite{Capstick:1986bm,Loring:2001kx}). Recent lattice gauge
calculations \cite{Edwards:2011jj} confirm these findings. However,
masses of resonances with positive and negative parities are often
similar, in striking disagreement with quark models and the results
on the lattice. A second problem of both, lattice calculations and
quark models, is the number of expected states which is considerably
larger than confirmed experimentally, a fact which is known as
problem of {\it missing resonances.} The number of expected states
is much reduced if it is assumed that two quarks form a quasi-stable
diquark \cite{Anselmino:1992vg}.
\paragraph{Diquarks:}
In the fourth resonance region, at about 2\,GeV, at least four
positive-parity nucleon resonances were found, even though some of
the solutions were ambiguous. In the most straightforward
interpretation, we use solution BnGa2011-01, and assign the four
states
\begin{equation}
N(1875){1/2^+},\qquad N(1915){3/2^+},\qquad N(1860){5/2^+},\qquad
N(1990){7/2^+}\nonumber
\end{equation} to a spin quartet of nucleon resonances
\cite{Anisovich:2011su}. This assignment excludes conventional
diquark models: A S-wave diquark is symmetric with respect to the
exchange of the two quarks, a third quark with even angular momentum
is symmetric with respect to the diquark, but the isospin wave
function of a nucleon resonance is of mixed-symmetry. Hence the
overall spin-flavor-spatial wave function is of mixed symmetry. With
an antisymmetric color wave function, the overall wave function has
no defined exchange symmetry and the Pauli principle would be
violated. With the assignment of the four resonances to a spin
quartet, the diquark hypothesis -  which freezes of one pair of
quarks into a quasi-stable S-wave flavor diquark - is ruled out as
explanation of the {\it missing resonance problem}.

\paragraph{Parity doublets:}

Ground-state baryons acquire their mass due to spontaneous breaking
of chiral symmetry. Thus, $N_{1/2^-}(1535)$ is much heavier than its
chiral partner, $N_{1/2^+}(940)$. At high excitation energies,
details of the chiral potential could be irrelevant, and chiral
symmetry could be restored \cite{Glozman:1999tk}. Then parity
doublets should occur. These have been observed, indeed. Some
previously known doublets are listed in Table~\ref{tab:chiral}. The
new resonances form parity doublets as well, in particular when
solution BnGa2011-02 is chosen.

\begin{table}[pt]
\bc
\begin{footnotesize}
\renewcommand{\arraystretch}{1.5}
\begin{tabular}{||c||c||c||c||}\hline\hline
$N(1650){1/2^-}$&$ N(1700){3/2^-}$&$
N(1675){5/2^-}$&\color{darkred}$
\Delta(2200){7/2^-}$\\
$N(1710){1/2^+}$&$ N(1720){3/2^+}$&$
N(1680){5/2^+}$&\color{darkred}$ \Delta(1950){7/2^+}$\\\hline\hline
\color{darkblue}$N(1895){1/2^-}$&$
\color{darkblue}N(1880){3/2^-}$&\color{darkblue}$
N(2060){5/2^-}$&\color{darkblue}$
N(2180){7/2^-}$\\
\color{darkblue}$N(1870){1/2^+}$&\color{darkblue}$
N(1890){3/2^+}$&\color{darkblue}$ N(2090){5/2^+}$&\color{darkblue}$
N(2105){7/2^+}$\\\hline\hline
\end{tabular}
\renewcommand{\arraystretch}{1.0}
\end{footnotesize}
\ec
\caption{\label{tab:chiral}Spin-parity doublets for $J=1/2$, $3/2$,
$5/2$ (first three boxes) and for $J=1/2,\cdots , 7/2$ (four boxes
in the lower part of the table) for nucleon and $\Delta$ resonances.
Meson and baryon resonances on the leading Regge trajectory like
$\Delta(1950){7/2^+}$ or $a_6(2450)$ have no mass-degenerate parity
partner; shown are $\Delta(1950){7/2^+}$ and $\Delta(2200)7/2^-$ but
other parity partners not degenerate in mass exist as well.}
\end{table}

\paragraph{AdS/QCD:}
There is, however, one caveat: both mesons and baryons on the
leading Regge trajectory have no mass-degenerate parity partner.
This selectivity in the formation of parity partners follows from
the AdS/QCD mass formula for $n\bar n$ mesons and for $\Delta$
baryons $M^2 = c + a (L+N)$ where $c=1/2$ for mesons and $3/2$ for
baryons, and where $a$ is the string constant \cite{Forkel:2007tz}.
For $\Delta$ and $N$ resonances, the formula can be extended
\cite{Forkel:2008un} to
\begin{eqnarray}
M^2 =& a\cdot (\mathtt{L+N}+3/2)-b\,\alpha_D \nonumber
\label{M_ADS}
\end{eqnarray}
which reproduces in a two-parameter fit the mass spectrum 2$\times$
better than quark models, Skyrme models (or LQCD).

\paragraph{Missing resonances:}
The search for {\it missing resonances} has been a driving force for
photoproduction experiments \cite{Klempt:2009pi}. The new resonances
suggest that the existing resonances fill all states expected in the
ground state and the first excitation shell. In the second shell,
two multiplets are completely filled, two are completely empty, in
the third shell, two are full, six are empty. There seems to be a
dynamical selection in the {\it missing resonances} which is not yet
understood \cite{Anisovich:2011sv}. One could anticipate that the
dynamical reason is due to a string-like nature of the quark-quark
interaction, and that the interpretation of the mass spectrum by
constituent quarks missed the increase in the mass of the string
when two quarks are dynamically separated in space. The consistency
between quark model and lattice predictions may teach us that quarks
leading to $m_\pi=400$\,MeV are still ``too static", and that
considerably lower quark masses have to be reached before
quantitative lattice predictions will reproduce or predict the
spectrum of light-quark baryon excitations.

\paragraph{\boldmath Do glueballs, hybrids, and multiquark states exist atop of $q\bar q$ and $qqq$ states?}
The density of nucleon excitations would increase further, and the
problem of the {\it missing resonances} aggravated, if baryonic
hybrids, baryons in which the gluon string is excited, would exist.
This is a more general problem in hadron spectroscopy. A large
variety of different species is predicted which may all be realized
independently. Of course, they can mixed but the number of predicted
states with the same quantum number increases when glueballs,
hybrids, and multiquark states exist on their own right. A well
known example are the two axial vector mesons with strangeness.
According to the quark model, one meson resonance is predicted along
with the $a_1(1260)$, the other one with $b_1(1230)$. They both have
$J^P=1^+$ but differ in $G$-parity. However, $G$-parity is not a
good quantum number for strange mesons, the two quark model states
$K_{1a}$ and $K_{1b}$ can therefore mix, and two resonances emerge
known as $K_{1}(1280)$ and $K_{1}(1400)$.

In baryon spectroscopy, five-quark $qqqq\bar q$ resonances could
exist in addition to the conventional three-quark states
\cite{Zou:2005xy}. Resonances generated dynamically by
channel-channel effects might lead to additional resonances. Here,
the question arises: is $N(1535)1/2^-$ a $qqq$ quark model state,
does its strong $\Lambda K$ coupling indicate a $qqqq\bar q$
resonance, or is it a $N\eta -\Lambda K$ coupled channel resonance
of dynamical origin? Likely, it is all of it, and the wave function
contains all components. But then, are there orthogonal states in
the spectrum?

Scalar mesons with different hadronic content are all predicted in
the 1 - 2\,GeV mass range. Do they mix and can they be observed
experimentally as five different states?

\bc
\begin{tabular}{ccccc}\hline
$q\bar q$ &  \color{darkblue}$q\bar qq\bar q$ & \color{darkbrown} $q\bar qg$ & \color{darkred}  $gg$ &  \color{darkgreen} $m_1m_2$ \\
mesons&  \color{darkblue}tetraquarks& \color{darkbrown} hybrids&
\color{darkred}  glueballs& \color{darkgreen} molecules\\\hline
\end{tabular}
\ec

\bc
\begin{tabular}{ccccc}\hline
$qqq$ & \color{darkblue}$qqqq\bar q$ &  \color{darkbrown}$qqqg$  && \color{darkgreen} $b_1m_2$\\
baryons&  \color{darkblue}pentaquarks& \color{darkbrown}
hybrids&\phantom{glueballs}& \color{darkgreen} molecules\\\hline
\end{tabular}
\ec

This question can best be answered in a discussion of low-lying
scalar nonets. Jaffe suggested that the lightest scalar mesons can
be interpreted as four-quark states with a pair of quarks in color
$\bar 3$ and a pair of antiquarks in color~$3$. In SU(3) this
condition can be fulfilled by nine quarks, and this is just the
number of light $q\bar q$ mesons in the scalar nonet. In SU(4),
adding charm, this pattern is different \cite{Klempt:2007cp}.
Restricting us to scalar mesons with open charm, there are now 10
configurations with a pair of quarks in color $\bar 3$ and a pair of
antiquarks in color $3$. Six of them can couple to $c\bar n$ or
$n\bar c$, four of them are flavor exotic. The latter configurations
have never been observed. We conjecture that the bindings forces
between quark and antiquark are essential to form mesons, that all
mesons must have a $q\bar q$ component. {\it This conjecture may
exclude the existence of spin-parity exotics as well and likely also
the existence of glueballs as additional states.} The conjecture
does not exclude that a scalar state has a sizable glueball
fraction, but supernumerosity is not expected.

For baryons, we then expect no additional states; some baryon
resonances are generated dynamically, but they need a $qqq$ seed to
exist. Flavor exotics are not expected.

\subsection*{Outlook} 
\begin{figure}[pt]
\bc
\includegraphics[width=0.90\textwidth,height=0.6\textwidth]{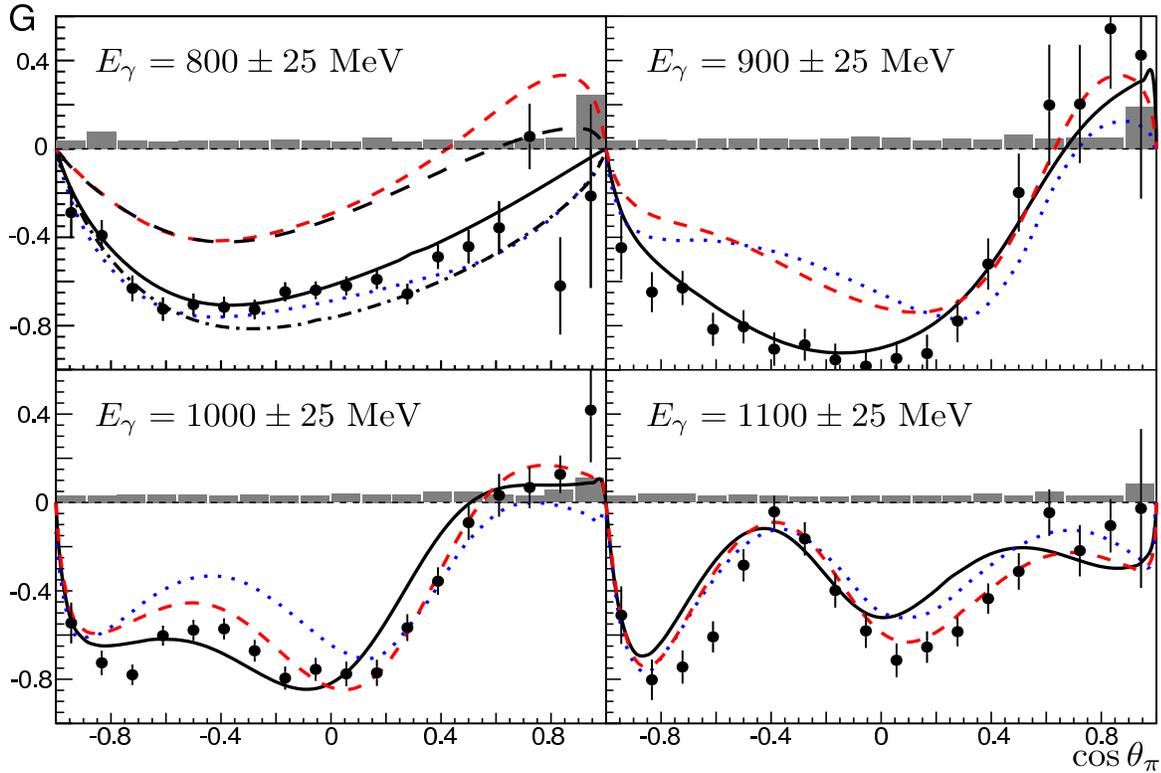}
\vspace{-5mm}\ec \caption{\label{pic:results}The polarization
observable $G$ as a function of $\cos\theta_{\pi}$ from $E_\gamma
=800$\,MeV up to $E_\gamma =1100$\,MeV. Systematic errors are shown
in gray bars. The curves represent predictions from different
partial wave analyses. Solid (black) curve: BnGa
\cite{Anisovich:2011fc}; dashed (red): SAID \cite{Dugger:2009pn};
long-dashed (black): BnGa with $E_{0^+}$ and $E_{2^-}$ amplitudes
from SAID; dotted (blue): MAID \cite{Drechsel:1998hk}; dashed-dotted
(black): BnGa with $E_{0^+}$ and $E_{2^-}$ amplitudes from MAID.
Gray area shows the systematic error due to interactions on nuclei
and uncertainty in the photon polarization. }
\end{figure}

Large data sets on photoproduction off
protons and off neutrons have been taken at Bonn, Jlab, and Mainz
with longitudinally and linearly polarized photons and with
longitudinal and transverse target polarizations. The aim is to
gather a complete data base which will define unambiguously the
nucleon excitation spectrum. The impact of double polarization data
on the spectrum can be seen from first data on the double
polarization variable $G$. The data on $G$ are shown in
Figure~\ref{pic:results} and compared to the predictions from the
SAID (SN11) \cite{Dugger:2009pn}, MAID \cite{Drechsel:1998hk}, and
BnGa \cite{Anisovich:2011fc} partial wave analyses. The results are
surprising. The predictions of the three PWA differ considerably
even in the region of the first negative-parity resonances. The data
follow the prediction from BnGa rather well while MAID
\cite{Drechsel:1998hk} and SAID \cite{Dugger:2009pn} are disfavored.
The reason for the discrepancies can be assigned to the different
photo-couplings of $N(1535)1/2^-$ and $N(1520)3/2^-$ \cite{Thiel}.

\paragraph{Acknowledgement} I would like
to thank the organizers for their kind invitation and hospitality,
and all members of the SFB/TR16 for fruitful discussions. The
SFB/TR16 is supported by DFG.

\end{document}